\documentclass[aps,prb,twocolumn,secnumarabic,amsmath,amssymb,superscriptaddress]{revtex4-1}

\usepackage{titlesec}

\usepackage{bm}
\usepackage{comment}

\usepackage{graphicx}
\usepackage{subfigure}
\usepackage{tabularx}
\usepackage{color}
\usepackage[colorlinks,bookmarks=false,citecolor=blue,linkcolor=red,urlcolor=blue]{hyperref}
\usepackage{verbatim}

\newcommand{\cuse}{Cu$_2$OSeO$_3$}

\definecolor{violet}{rgb}{0.8,0.2,0.6}

\def\ua{\uparrow}
\def\da{\downarrow}
\def\uua{\Uparrow}
\def\dda{\Downarrow}
\def\be{\begin{equation}}
\def\ee{\end{equation}}
\def\bea{\begin{eqnarray}}
\def\eea{\end{eqnarray}}

\def\vec{\mathbf}

\def\mc{\mathcal}

\usepackage{times}
\begin{document}

\title{The quantum origins of skyrmions and half-skyrmions in Cu$_2$OSeO$_3$}

\author{Oleg Janson}
\affiliation{Max Planck Institute for Chemical Physics of Solids, Dresden, D-01087, Germany}
\affiliation{National Institute of Chemical Physics and Biophysics, Tallinn, EE-12618, Estonia}

\author{Ioannis Rousochatzakis}
\affiliation{Leibniz Institute for Solid State and Materials Research, IFW Dresden, D-01069, Germany}

\author{Alexander A. Tsirlin}
\affiliation{Max Planck Institute for Chemical Physics of Solids, Dresden, D-01087, Germany}
\affiliation{National Institute of Chemical Physics and Biophysics, Tallinn, EE-12618, Estonia}

\author{Marilena Belesi}
\affiliation{Leibniz Institute for Solid State and Materials Research, IFW Dresden, D-01069, Germany}

\author{Andrei A. Leonov}
\affiliation{Leibniz Institute for Solid State and Materials Research, IFW Dresden, D-01069, Germany}

\author{Ulrich K. R\"{o}{\ss}ler}
\affiliation{Leibniz Institute for Solid State and Materials Research, IFW Dresden, D-01069, Germany}

\author{Jeroen van den Brink}
\affiliation{Leibniz Institute for Solid State and Materials Research, IFW Dresden, D-01069, Germany}
\affiliation{Department of Physics, TU Dresden, D-01062 Dresden, Germany}

\author{Helge Rosner}
\affiliation{Max Planck Institute for Chemical Physics of Solids, Dresden, D-01087, Germany}

\begin{abstract}
The Skyrme-particle, the {\it skyrmion}, was introduced over half a century ago
and used to construct field theories for dense nuclear matter.\cite{skyrme1962,
Witten1983} But with skyrmions being mathematical objects\,---\,special types of
topological solitons\,---\,they can emerge in much broader
contexts.\cite{bogdanov1989, bogdanov1994, roessler2006} Recently skyrmions
were observed in helimagnets,\cite{muehlbauer2009,tonomura2012,yu2010,Cu2OSeO3_skyrmions,yu2012} 
forming nanoscale spin-textures that hold promise as information carriers.\cite{jonietz2010,fert2013}
Extending over length-scales much larger than the inter-atomic spacing, these
skyrmions behave as large, classical objects, yet deep inside they are of
quantum origin. Penetrating into their microscopic roots requires a multi-scale
approach, spanning the full quantum to classical domain. By exploiting a
natural separation of exchange energy scales, we achieve this for the first
time in the skyrmionic Mott insulator Cu$_2$OSeO$_3$. Atomistic {\it ab initio}
calculations reveal that its magnetic building blocks are strongly fluctuating
Cu$_4$ tetrahedra.  These spawn a continuum theory with a skyrmionic texture
that agrees well with reported experiments. It also brings to light a decay of
skyrmions into {\it half-skyrmions} in a specific temperature and magnetic
field range.  The theoretical multiscale approach explains the strong
renormalization of the local moments and predicts further fingerprints of the
quantum origin of magnetic skyrmions that can be observed in \cuse, like weakly
dispersive high-energy excitations associated with the Cu$_4$ tetrahedra, a
weak antiferromagnetic modulation of the primary ferrimagnetic order, and a
fractionalized skyrmion phase.
\end{abstract}

\maketitle

\begin{figure*}[!t]
\begin{center}
\includegraphics[width=0.7\linewidth]{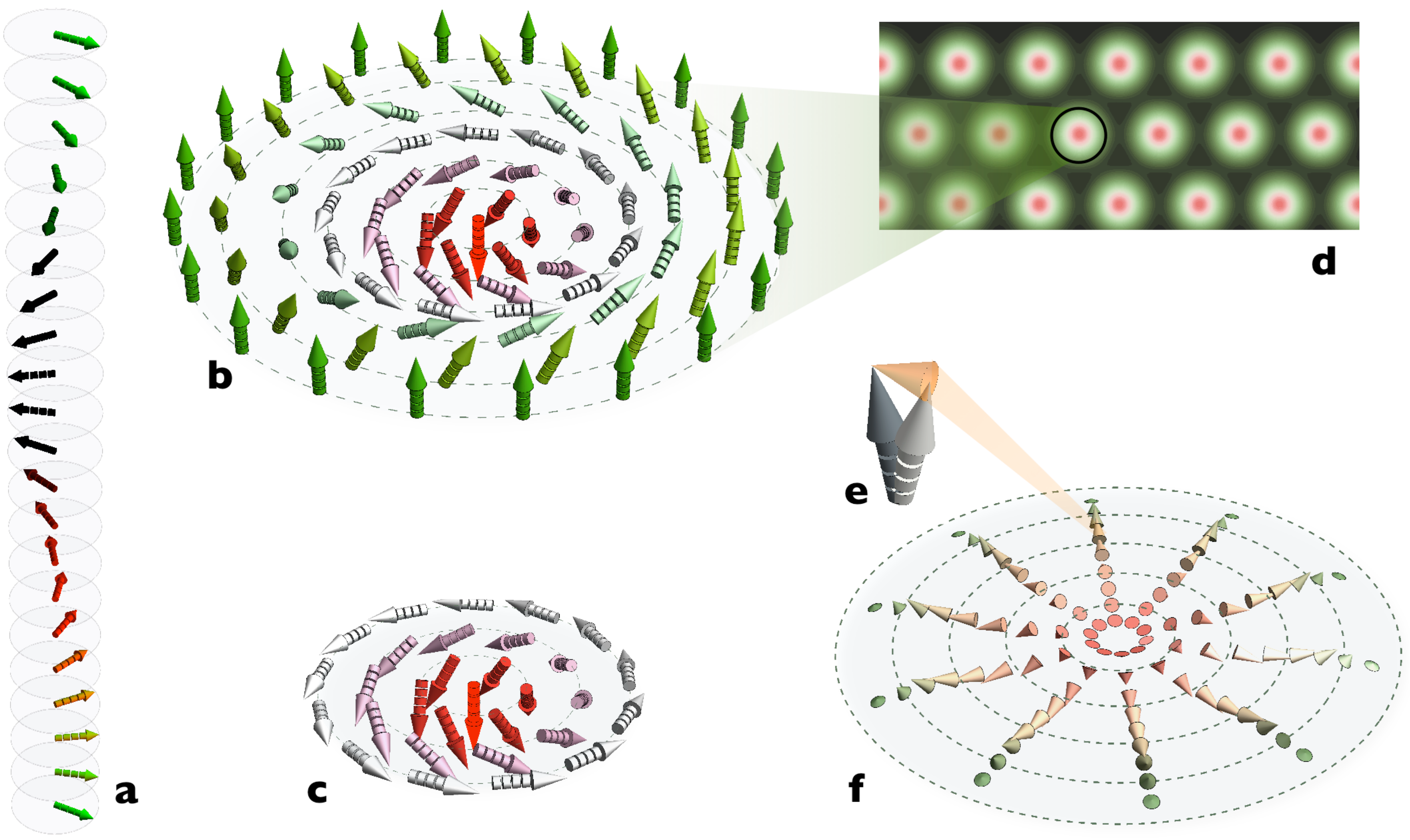}\end{center}
\caption{
Besides flat helices ({\bf a}), chiral helimagnets like \cuse, manifest
radially symmetric topological solitons, like skyrmions ({\bf b}) or
half-skyrmions ({\bf c}), where the local order parameter (sectioned arrows)
forms a double-twisted core, tracing out the whole ({\bf b}) or half ({\bf c})
of the Bloch sphere.  {\bf d}, Parallel skyrmions can form densely packed
lattices in two spatial dimensions.  {\bf e}, Quantitative first-principles
calculations predict that the ferrimagnetic order in \cuse\ is locally altered
by the multi-sublattice structure.  Such a canted arrangement is usually called
a weak antiferromagnetic order.  {\bf f}, The skyrmion texture is locally
composed of these three-dimensional canted spin patterns. Thus, the weak
antiferromagnetic order itself is modulated along with the primary
ferrimagnetic twisting shown in {\bf b}.\label{fig:skyrmions} }
\end{figure*}

Skyrmionic spin textures in magnetic materials correspond to magnetic
topological solitons as depicted in Fig.~\ref{fig:skyrmions}. They were first
observed in the non-centrosymmetric B20 helimagnets
MnSi,\cite{muehlbauer2009,tonomura2012}, FeGe,\cite{yu2012} and
Fe$_{0.5}$Co$_{0.5}$Si.\cite{yu2010}  These skyrmionic textures are encountered
also in a completely different branch of physics: in the theoretical
description of nuclear matter.\cite{skyrme1962,Witten1983} In this setting the
skyrmions are of course not related to magnetic degrees of freedom, but rather
to particles emerging from cold hadron vector fields at densities a few times
that of ordinary nuclear matter. This is the density range relevant for compact
astronomical objects such as neutron stars.\cite{Ouyed1999,Luckock1986,Lee2010}
The perhaps perplexing connection between these two seemingly disparate fields
of physics is borne out of the underlying mathematical
structures.\cite{bogdanov1989, bogdanov1994,
roessler2006,Lee2010,Ouyed1999,Luckock1986,Castillejo1989,Kugler1989} The
physical phenomena in the two different settings are both governed by an
emerging set of differential equations with topological solitonic solutions:
the skyrmions found first by Skyrme in the 1960s.\cite{skyrme1962} 

In this context we investigate the formation and microscopic origin of the
observed magnetic skyrmions in helimagnets (Fig.~\ref{fig:skyrmions}). These
skyrmions are large objects compared to the atomic length-scale: they are about
three orders of magnitude larger in size than the inter-atomic lattice spacing.
Understanding the origin of these nanometer-scale skyrmions therefore requires
a multi-scale approach. In the above mentioned B20 helimagnets such is however
not viable because all these materials are metallic. The metallicity causes
low-energy, delocalized electronic and magnetic degrees of freedom to mix so
that they intrinsically involve multiple energy and spatial scales, which
renders a multi-scale approach presently intractable.

\begin{figure*}[!t]
\includegraphics[width=\linewidth]{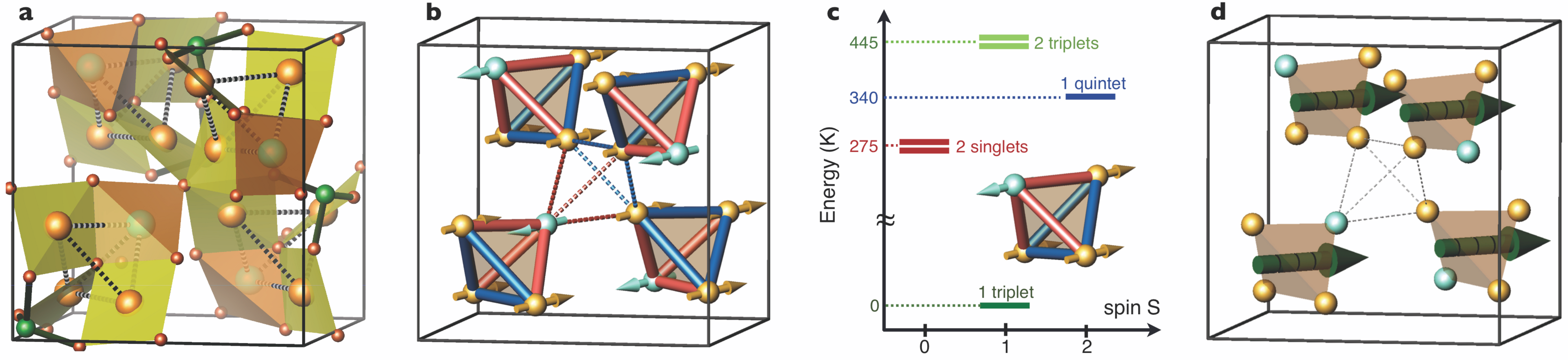}
\caption{\label{fig:str_model} 
Multiscale modeling of Cu$_2$OSeO$_3$. {\bf a}, The crystal structure is shaped
by Cu(1)O$_4$ plaquettes (yellow) and Cu(2)O$_5$ bipyramids (orange), and
covalent Se--O bonds (thick lines), forming a sparse three-dimensional lattice.
This lattice can be tiled into tetrahedra (dashed lines), each composed of one
Cu(1) and three Cu(2) polyhedra. 
\textbf{b}, The magnetic Cu$^{2+}$ ions form a distorted pyrochlore lattice, a
network of corner-shared tetrahedra. DFT calculations evidence the presence of
both types of magnetic interactions\,---\,antiferromagnetic (red) and
ferromagnetic (blue), in agreement with experimental magnetic structure
(arrows).  The strength of a certain coupling is indicated by the thickness of
the respective line. The strongest couplings are found within the tetrahedra
(shaded), while the couplings between the tetrahedra (dashed lines) are
substantially weaker. 
\textbf{c}, A quantum mechanical treatment of a single tetrahedron yields a
magnetic spin $S\!=\!1$ ground state, separated from the lowest lying
excitation by $\sim $275\,K. Due to this large energy scale, the tetrahedra
behave as rigid $S_t\!=\!1$ entities at low temperatures.  
\textbf{d}, The effective $S_t\!=\!1$ entities reside at the vertices
of a {\it trillium} lattice, exactly like the Mn ions in MnSi. Their mutual
effective exchange couplings are all ferromagnetic (see text). The
quantum-mechanical nature of the effective $S_t\!=\!1$ moments is
indicated by sectioned arrows.}
\end{figure*}

This is very different in the recently discovered skyrmionic material \cuse, a
large band-gap Mott insulator (Fig.~\ref{fig:str_model}). The band gap enforces
a natural separation between electronic and magnetic energy scales.  \cuse\ is
actually the first example of an insulating material displaying the chiral
helimagnetism that is desired for skyrmion formation while sharing the
non-centrosymmetric cubic space group $P2_13$ of the metallic B20 phases, but
with a unit cell that is much more complex, containing 16 Cu atoms. Due to the
presence of a magnetoelectric coupling\cite{Cu2OSeO3_skyrmions, Cu2OSeO3_MEH}
its skyrmions can be manipulated by an electric field,\cite{White} which is in
principle very energy efficient as this avoids losses due to joule heating. 

A multi-scale approach to elucidate the quantum origin of the skyrmion textures
in \cuse\ has to start from a calculation of magnetic interactions at the
atomic level. In a \cuse\ crystal, the magnetic Cu$^{2+}$ ions make up a 3D
network of corner-sharing tetrahedra (Fig.~\ref{fig:str_model}, b) with two
inequivalent Cu sites, Cu(1) and Cu(2) that are inside Cu(1)O$_5$ bi-pyramids
and distorted Cu(2)O$_4$ plaquettes, respectively.\cite{Bos08,Cu2OSeO3_NMR}
Each tetrahedron contains Cu(1) and Cu(2) in a ratio of 1:3. The resulting net
of magnetic Cu ions in \cuse\ thus has a structure that is rather different
from the previously mentioned metallic B20 helimagnets such as MnSi, in which
the magnetic Mn ions constitute instead a three dimensional corner-sharing net
of triangles, commonly referred to as the {\it trillium} lattice. The more
complex crystal structure of \cuse\ leads to five inequivalent superexchange
coupling constants $J_{ij}$ between neighboring $S\!=\!1/2$ copper spins $i$
and $j$ and also five different Dzyaloshinskii-Moriya (DM) vectors
$\vec{D}_{ij}$ in the microscopic magnetic Hamiltonian 
\be
\mc{H}=\sum_{i>j} J_{ij} \vec{S}_i\cdot\vec{S}_j  + \vec{D}_{ij}\cdot \vec{S}_i \times \vec{S}_j~,
\ee
where $\vec{S}_i$ denotes the Cu quantum-spin at site $i$. We have determined
these coupling constants by means of an extended set of {\it ab initio} density
functional based electronic structure calculations.  The obtained values were
cross-checked by calculating the magnetic $T_C$ and the temperature dependence
of both the magnetization and magnetic susceptibility by means of large scale
Quantum Monte Carlo (QMC) simulations.  These simulations agreeing very well
with the measurements inspire  further confidence in the accuracy of the values
calculated from first principles.

Having fixed the microscopic coupling constants, we proceed by establishing a
hierarchy of magnetic energy scales. We first observe that the calculated $J$'s
between Cu(1) and Cu(2) are antiferromagnetic (AFM) and between Cu(2) and Cu(2)
ferromagnetic (FM). A more detailed examination of the magnetic energy scales
reveals a striking difference between two groups of exchange couplings,
splitting the system into two kinds of Cu$_4$ tetrahedra: one with strong
($|J_s|\!\sim\!130\!-\!170$\,K) and the other with weak
($|J_w|\!\sim\!30\!-\!50$\,K) superexchange couplings. The DM terms are in turn
much smaller than the exchange couplings, $|\vec{D}_{ij}| \!\ll\! |J|$. 

The four $S\!=\!1/2$ spins of a Cu$_4$ tetrahedron with strong superexchange
interactions can couple together to form either a total singlet, triplet or
quintet state (with total spin $S_t\!=\!0$, 1 or 2, respectively). The
tetrahedron having 3 AFM and 3 FM exchange couplings, renders the ground state
(GS) a total $S_t\!=\!1$ triplet, see Fig.~\ref{fig:str_model} (b-c). The
triplet GS is separated from the other spin-multiplets by a large energy gap of
$\sim$275\,K.  An important point is that the Cu$_4$ tetrahedron triplet
wavefunction is not the classical (tensor product) state
\mbox{$|\!\ua\ua\ua\dda\rangle$} (where the double arrow labels the Cu(1) site
in the tetrahedron) but rather a coherent quantum superposition of four
classical states
\be\label{eq:triplet}
|S_t\!=\!1,\!M\!=\!1\rangle\!=\!\frac{1}{\sqrt{12}} \Big(
3|\!\ua\ua\ua\dda\rangle-|\!\da\ua\ua\uua\rangle-|\!\ua\da\ua\uua\rangle-|\!\ua\ua\da\uua\rangle \nonumber
\Big),
\ee
with $M$ labeling the three orthogonal triplet states with $M$\,=\,$-1$,\,0,\,1 (for
brevity only the $M\!=\!1$ wavefunction is given above). Although these are not
the exact tetrahedron basis states due to the presence of a small
triplet-quintet mixing, this representation is qualitatively
correct. This effective spin wavefunction is in full agreement with the
experimental observation of a locally ferrimagnetic order
parameter.\cite{Bos08,Cu2OSeO3_NMR} The quantum fluctuations ingrained into
these triplet wavefunctions, however, give rise to a substantial reduction of
the local moments, providing a natural explanation for the origin of the small
moments observed experimentally.\cite{Bos08} As opposed to transversal spin
fluctuations arising from spin waves (which are expected to be small in the
present case owing to the dimensionality, the ferrimagnetic nature of the order
parameter and the absence of frustration), these local quantum fluctuations are
longitudinal in character and hence directly affect the effective magnitude of
the spin. This picture is confirmed by a lattice QMC simulation for the full
model of Cu $S\!=\!1/2$ spins.

This establishes Cu$_4$ tetrahedra carrying magnetic triplets as building
blocks in \cuse\ at the next step of the multi-scale approach. Within this
abstraction, each of these tetrahedra can be contracted to a single lattice
point. The resulting structure turns out to consist of corner-shared triangles
which together constitute a trillium lattice, which is
precisely the same lattice that is formed by the Mn atoms in the B20 structure
of MnSi and the Fe atoms in FeGe. This establishes a very close analogy between
Mott insulating \cuse\ and these well-known metallic helimagnets, despite the
fundamental differences in electronic structure. However in \cuse, the
effective triplet interactions can be derived relying on rigorous microscopic
results. At this point both the weaker superexchange couplings $J_w$ and the DM
interaction $D_{ij}$ become crucial. A straightforward perturbative calculation
reveals that their net effect is a weak FM interaction between nearest-neighbor
(NN) and next-nearest-neighbor (NNN) $S_t\!=\!1$ spins, with an effective
exchange coupling of about $-8$\,K, which reflects the tendency of the system
towards FM ordering. This drastic reduction of the energy scale in the
effective model is caused by the renormalization of the local spin lengths and
the strong quantum correlations inside the strongly coupled tetrahedra. Not
only an exchange interaction, but also a DM coupling between NN and NNN
$S_t\!=\!1$ spins emerges. This is crucial because in the GS of a single strong
tetrahedron all diagonal matrix elements of the DM couplings {\it within} the
tetrahedron vanish by symmetry. The twisting mechanism that causes chiral
helimagnetism in \cuse, therefore originates from the effective DM couplings
{\it between} the strong Cu$_4$ tetrahedra: these will be the root cause for
skyrmions to emerge.

Having established the effective trillium lattice model of \cuse, we now
proceed to the long wavelength magnetic continuum theory that governs the
skyrmion formation in \cuse\ on the mesoscopic scale. The resulting continuum
equations involve two magnetic constants, $\mc{J}$ and
$\mc{D}$, which from a direct calculation are evaluated to be
$\mc{J}\!\simeq\!15$\,K and  $\mc{D}\!\simeq\!3$\,K. With the characteristic
period $\Lambda$ of the double-twisted skyrmion structures being $2\pi/Q$, with
wavenumber $Q=\frac{\mc{D}}{2\mc{J}}\frac{1}{\text{a}}$ and lattice constant
$\text{a}$, the calculated magnetic constants result in a helix period of
$\Lambda\!\simeq\!20$\,nm which has the correct order of magnitude compared
with the experimentally measured value\cite{Cu2OSeO3_skyrmions} of 50\,nm.
Besides this agreement with basic experimental observations, our multi-scale
description also provides two essential predictions.  Firstly, a very distinct
set of weakly dispersive, high-energy intra-tetrahedra excitation modes should
appear. Secondly, a specific antiferromagnetic superstructure emerges that is
the dual counterpart of the weak {\it ferro}magnetism present in chiral
acentric bipartite antiferromagnets.\cite{bogdanov2002} Both these effects
can immediately be tested experimentally, for instance by neutron scattering.

\begin{figure*}[!t]
\includegraphics[width=\linewidth]{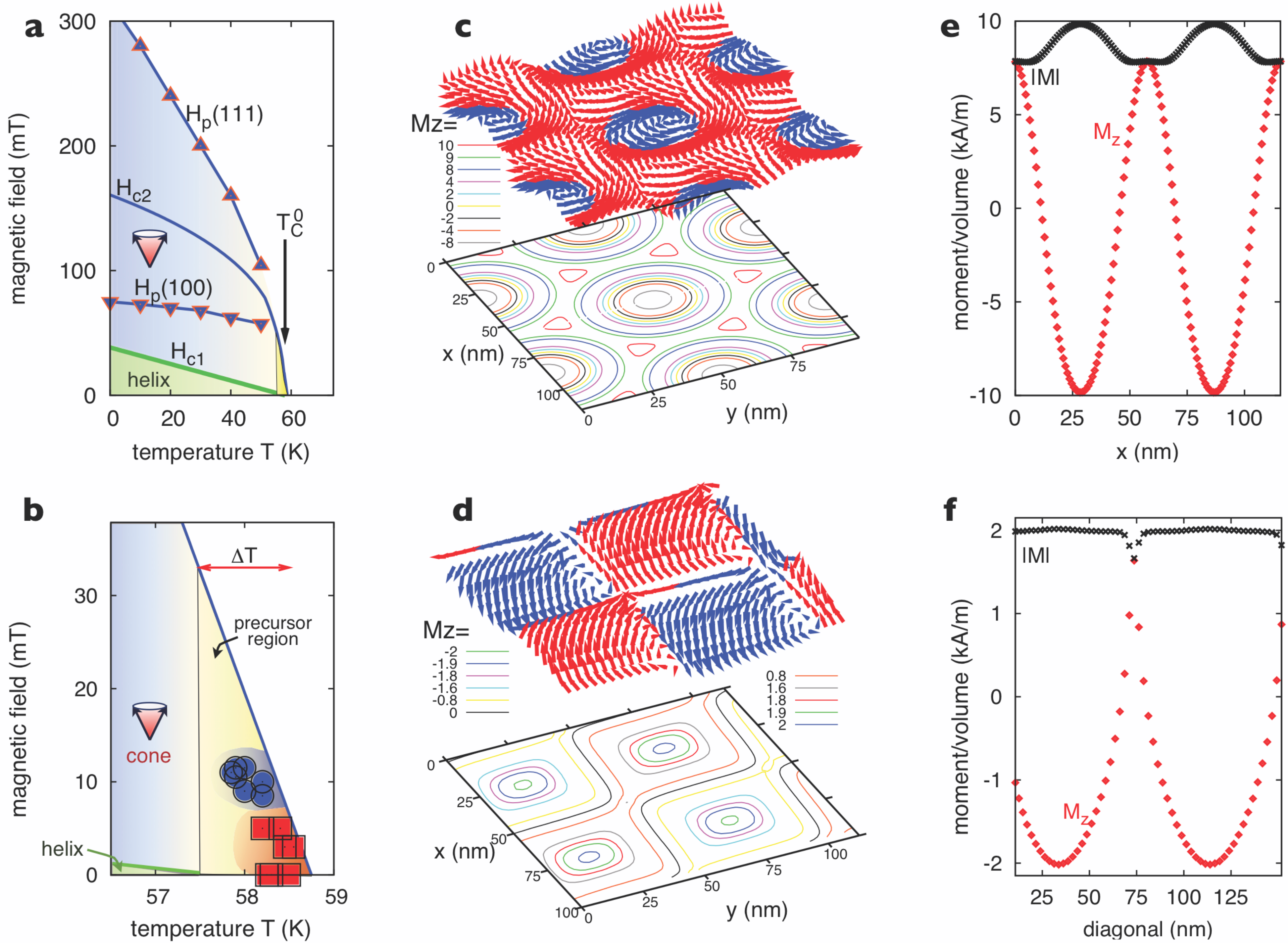}
\caption{\label{fig:continuum_model} 
Phase diagram of \cuse\ from Landau-Ginzburg continuum description.
{\bf a}, $H_{c1}$ denotes the reorientation transition of the helices
into the conical state.  Without additional anisotropies, the cone angle closes
continuously at $H_{c2}$ and the system reaches the ferrimagnetic plateau
phase. A cubic anisotropy makes $H_{c2}$ direction depending.  $H_p(111)$
denotes the continuous transition for fields along $[$111$]$.  For fields along
$[$100$]$ direction, the conical helix collapses by a first-order process at
$H_p(100)$.  {\bf b}, The high-temperature phase diagram has a narrow precursor
region where skyrmionic phases are found numerically for two-dimensional
models.  Blue circles show the region of stable densely packed $-\pi$-skyrmion
lattices (sketched in {\bf c} with contour plot for component $M_z$ and
corresponding profiles across nearest neighbour skyrmions {\bf e}); red squares
mark the region of stable $\pm \pi/2$-skyrmion lattices (sketched in {\bf d}
with profiles along a next-nearest-neighbour diagonal {\bf f}).
}
\end{figure*}

Having quantified in detail the microscopic couplings responsible for the
chirally twisted spin order in \cuse, its magnetic phase diagram can be
assessed. In the continuum description of \cuse,  the Dzyaloshinskii model of
chiral cubic ferromagnets, we can merge the {\it ab initio} parameters and the
exchange stiffness ${\cal A}$ (obtained from QMC calculations) to determine
quantitatively the thermodynamic (Landau) potential. The value of the weakest
primary coupling, the DM interaction, is fixed by the experimentally observed
helix period.  With this approach both the magnetic field and
temperature scale are fully determined. 

In this framework, one can calculate the critical field $H_{c2}$ for the
continuous transition from the conical helix state into the field-enforced
ferrimagnetic collinear state. We find 80~mT at 50~K, in very good agreement
with experimental data.  We also determined the temperature window for the {\it
precursor region} at around the magnetic ordering temperature where
meso-phases, that are potentially of skyrmionic character, can be formed and
find $\Delta T\!\simeq\!1$~K . In this temperature interval, the skyrmionic
cores are energetically favorable compared to one-dimensional helix
solutions.\cite{roessler2006} The computed range is in agreement with the
interval of about 2~K in which the so-called A-phases appear under magnetic
field in \cuse\ crystals.\cite{Cu2OSeO3_NS,Cu2OSeO3_ND_MH_single_cryst} From
symmetry considerations it is immediately clear that the dominant anisotropy in
\cuse\ is cubic with a coefficient $K_{c1}\!>\!0$, which can stem from the
magnetoelectric effect and the dielectric polarizability of \cuse.
The experimental value for the magnetic field at which the conical
helix closes and becomes a field-enforced saturated state fixes this anisotropy
at  $K_{c1}\,M_{\text{sat}}^4 = 1.2 \cdot 10^{4}$~J/m$^3$ (corresponding to
3.3~$\mu$eV/Cu atom).

With all this in place, we can fully determine the equilibrium solutions and
thereby the phase diagram (Fig.~\ref{fig:continuum_model}~a and b). In the
precursor region, we find as equilibrium states two competing skyrmionic phases
(Fig.~\ref{fig:continuum_model}~c to f). The first one
(Fig.~\ref{fig:continuum_model}~c and e) is the standard field-driven
``$-\pi$''-skyrmion phase of Fig.~\ref{fig:skyrmions}(b) with the radial
skyrmions ordered in a hexagonal lattice.\cite{bogdanov1989,bogdanov1994} The
other one (Fig.~\ref{fig:continuum_model} d and f) is the ``$\pi/2$''-skyrmion\cite{bogdanov1994,Wilhelm2012} 
state of Fig.~\ref{fig:skyrmions}(c), which actually is the stable state at
zero and low fields, because the fractionalization of skyrmions into
half-skyrmions yields a higher packing density of the energetically
advantageous skyrmionic cores.  
The fractionalized skyrmion textures contain defects like hedgehogs, narrow line or wall
defects, where the magnetic order parameter passes through zero, leading to 
a broad distribution of local moments that can be discernible by local probes
such as $\mu$SR and NMR, or neutron diffraction methods. 
The emergence of the half-skyrmion phase in the vicinity of the $\pi$-skyrmion lattice
of \cuse\ opens a new venue to study properties of textures 
with split topological units in experiment. 
Observations of these defect-ridden topological phases, 
together with the predicted weakly antiferromagnetic indentations of 
the ferrimagnetic order, the strong fluctuations of the local moments, 
and the weakly dispersive high-energy magnetic excitations, that are
associated with the rigidly coupled spins of tetrahedra, allow to probe 
the quantum origin of the magnetic skyrmions in experiments on \cuse. 

\begin{acknowledgments}
We acknowledge fruitful discussions with J.-P.~Ansermet, A.~N.~Bogdanov,
V.~A.~Chizhikov, V.~E.~Dmitrienko, and Y. Onose. IR was supported by the
Deutsche Forschungsgemeinschaft (DFG) under the Emmy-Noether program. OJ and AT
were partly supported by the Mobilitas program of the ESF, grant numbers MJD447
and MTT77, respectively.
\end{acknowledgments}


\begin{thebibliography}{22}%
\makeatletter
\providecommand \@ifxundefined [1]{%
 \@ifx{#1\undefined}
}%
\providecommand \@ifnum [1]{%
 \ifnum #1\expandafter \@firstoftwo
 \else \expandafter \@secondoftwo
 \fi
}%
\providecommand \@ifx [1]{%
 \ifx #1\expandafter \@firstoftwo
 \else \expandafter \@secondoftwo
 \fi
}%
\providecommand \natexlab [1]{#1}%
\providecommand \enquote  [1]{``#1''}%
\providecommand \bibnamefont  [1]{#1}%
\providecommand \bibfnamefont [1]{#1}%
\providecommand \citenamefont [1]{#1}%
\providecommand \href@noop [0]{\@secondoftwo}%
\providecommand \href [0]{\begingroup \@sanitize@url \@href}%
\providecommand \@href[1]{\@@startlink{#1}\@@href}%
\providecommand \@@href[1]{\endgroup#1\@@endlink}%
\providecommand \@sanitize@url [0]{\catcode `\\12\catcode `\$12\catcode
  `\&12\catcode `\#12\catcode `\^12\catcode `\_12\catcode `\%12\relax}%
\providecommand \@@startlink[1]{}%
\providecommand \@@endlink[0]{}%
\providecommand \url  [0]{\begingroup\@sanitize@url \@url }%
\providecommand \@url [1]{\endgroup\@href {#1}{\urlprefix }}%
\providecommand \urlprefix  [0]{URL }%
\providecommand \Eprint [0]{\href }%
\providecommand \doibase [0]{http://dx.doi.org/}%
\providecommand \selectlanguage [0]{\@gobble}%
\providecommand \bibinfo  [0]{\@secondoftwo}%
\providecommand \bibfield  [0]{\@secondoftwo}%
\providecommand \translation [1]{[#1]}%
\providecommand \BibitemOpen [0]{}%
\providecommand \bibitemStop [0]{}%
\providecommand \bibitemNoStop [0]{.\EOS\space}%
\providecommand \EOS [0]{\spacefactor3000\relax}%
\providecommand \BibitemShut  [1]{\csname bibitem#1\endcsname}%
\let\auto@bib@innerbib\@empty
\bibitem [{\citenamefont {Skyrme}(1962)}]{skyrme1962}%
  \BibitemOpen
  \bibfield  {author} {\bibinfo {author} {\bibfnamefont {T.}~\bibnamefont
  {Skyrme}},\ }\href {\doibase 10.1016/0029-5582(62)90775-7} {\bibfield
  {journal} {\bibinfo  {journal} {Nucl. Phys.}\ }\textbf {\bibinfo {volume}
  {31}},\ \bibinfo {pages} {556} (\bibinfo {year} {1962})}\BibitemShut
  {NoStop}%
\bibitem [{\citenamefont {Witten}(1983)}]{Witten1983}%
  \BibitemOpen
  \bibfield  {author} {\bibinfo {author} {\bibfnamefont {E.}~\bibnamefont
  {Witten}},\ }\href@noop {} {\bibfield  {journal} {\bibinfo  {journal} {Nucl.
  Phys. B}\ }\textbf {\bibinfo {volume} {223}},\ \bibinfo {pages} {422 }
  (\bibinfo {year} {1983})}\BibitemShut {NoStop}%
\bibitem [{\citenamefont {Bogdanov}\ and\ \citenamefont
  {Yablonskii}(1989)}]{bogdanov1989}%
  \BibitemOpen
  \bibfield  {author} {\bibinfo {author} {\bibfnamefont {A.~N.}\ \bibnamefont
  {Bogdanov}}\ and\ \bibinfo {author} {\bibfnamefont {D.~A.}\ \bibnamefont
  {Yablonskii}},\ }\href@noop {} {\bibfield  {journal} {\bibinfo  {journal}
  {Zh. Eksp. Teor. Fiz.}\ }\textbf {\bibinfo {volume} {95}},\ \bibinfo {pages}
  {178} (\bibinfo {year} {1989})}\BibitemShut {NoStop}%
\bibitem [{\citenamefont {Bogdanov}\ and\ \citenamefont
  {Hubert}(1994)}]{bogdanov1994}%
  \BibitemOpen
  \bibfield  {author} {\bibinfo {author} {\bibfnamefont {A.}~\bibnamefont
  {Bogdanov}}\ and\ \bibinfo {author} {\bibfnamefont {A.}~\bibnamefont
  {Hubert}},\ }\href {\doibase 10.1016/0304-8853(94)90046-9} {\bibfield
  {journal} {\bibinfo  {journal} {J. Magn. Magn. Mater.}\ }\textbf {\bibinfo
  {volume} {138}},\ \bibinfo {pages} {255} (\bibinfo {year}
  {1994})}\BibitemShut {NoStop}%
\bibitem [{\citenamefont {R{\"{o}}{\ss}ler}\ \emph {et~al.}(2006)\citenamefont
  {R{\"{o}}{\ss}ler}, \citenamefont {Bogdanov},\ and\ \citenamefont
  {Pfleiderer}}]{roessler2006}%
  \BibitemOpen
  \bibfield  {author} {\bibinfo {author} {\bibfnamefont {U.~K.}\ \bibnamefont
  {R{\"{o}}{\ss}ler}}, \bibinfo {author} {\bibfnamefont {A.~N.}\ \bibnamefont
  {Bogdanov}}, \ and\ \bibinfo {author} {\bibfnamefont {C.}~\bibnamefont
  {Pfleiderer}},\ }\href {\doibase 10.1038/nature05056} {\bibfield  {journal}
  {\bibinfo  {journal} {Nature}\ }\textbf {\bibinfo {volume} {442}},\ \bibinfo
  {pages} {797} (\bibinfo {year} {2006})}\BibitemShut {NoStop}%
\bibitem [{\citenamefont {M{\"{u}}hlbauer}\ \emph {et~al.}(2009)\citenamefont
  {M{\"{u}}hlbauer}, \citenamefont {Binz}, \citenamefont {Jonietz},
  \citenamefont {Pfleiderer}, \citenamefont {Rosch}, \citenamefont {Neubauer},
  \citenamefont {Georgii},\ and\ \citenamefont {B{\"{o}}ni}}]{muehlbauer2009}%
  \BibitemOpen
  \bibfield  {author} {\bibinfo {author} {\bibfnamefont {S.}~\bibnamefont
  {M{\"{u}}hlbauer}}, \bibinfo {author} {\bibfnamefont {B.}~\bibnamefont
  {Binz}}, \bibinfo {author} {\bibfnamefont {F.}~\bibnamefont {Jonietz}},
  \bibinfo {author} {\bibfnamefont {C.}~\bibnamefont {Pfleiderer}}, \bibinfo
  {author} {\bibfnamefont {A.}~\bibnamefont {Rosch}}, \bibinfo {author}
  {\bibfnamefont {A.}~\bibnamefont {Neubauer}}, \bibinfo {author}
  {\bibfnamefont {R.}~\bibnamefont {Georgii}}, \ and\ \bibinfo {author}
  {\bibfnamefont {P.}~\bibnamefont {B{\"{o}}ni}},\ }\href {\doibase
  10.1126/science.1166767} {\bibfield  {journal} {\bibinfo  {journal}
  {Science}\ }\textbf {\bibinfo {volume} {323}},\ \bibinfo {pages} {915}
  (\bibinfo {year} {2009})}\BibitemShut {NoStop}%
\bibitem [{\citenamefont {Tonomura}\ \emph {et~al.}(2012)\citenamefont
  {Tonomura}, \citenamefont {Yu}, \citenamefont {Yanagisawa}, \citenamefont
  {Matsuda}, \citenamefont {Onose}, \citenamefont {Kanazawa}, \citenamefont
  {Park},\ and\ \citenamefont {Tokura}}]{tonomura2012}%
  \BibitemOpen
  \bibfield  {author} {\bibinfo {author} {\bibfnamefont {A.}~\bibnamefont
  {Tonomura}}, \bibinfo {author} {\bibfnamefont {X.}~\bibnamefont {Yu}},
  \bibinfo {author} {\bibfnamefont {K.}~\bibnamefont {Yanagisawa}}, \bibinfo
  {author} {\bibfnamefont {T.}~\bibnamefont {Matsuda}}, \bibinfo {author}
  {\bibfnamefont {Y.}~\bibnamefont {Onose}}, \bibinfo {author} {\bibfnamefont
  {N.}~\bibnamefont {Kanazawa}}, \bibinfo {author} {\bibfnamefont {H.~S.}\
  \bibnamefont {Park}}, \ and\ \bibinfo {author} {\bibfnamefont
  {Y.}~\bibnamefont {Tokura}},\ }\href {\doibase 10.1021/nl300073m} {\bibfield
  {journal} {\bibinfo  {journal} {Nano Letters}\ }\textbf {\bibinfo {volume}
  {12}},\ \bibinfo {pages} {1673} (\bibinfo {year} {2012})}\BibitemShut
  {NoStop}%
\bibitem [{\citenamefont {Yu}\ \emph {et~al.}(2010)\citenamefont {Yu},
  \citenamefont {Onose}, \citenamefont {Kanazawa}, \citenamefont {Park},
  \citenamefont {Han}, \citenamefont {Matsui}, \citenamefont {Nagaosa},\ and\
  \citenamefont {Tokura}}]{yu2010}%
  \BibitemOpen
  \bibfield  {author} {\bibinfo {author} {\bibfnamefont {X.~Z.}\ \bibnamefont
  {Yu}}, \bibinfo {author} {\bibfnamefont {Y.}~\bibnamefont {Onose}}, \bibinfo
  {author} {\bibfnamefont {N.}~\bibnamefont {Kanazawa}}, \bibinfo {author}
  {\bibfnamefont {J.~H.}\ \bibnamefont {Park}}, \bibinfo {author}
  {\bibfnamefont {J.~H.}\ \bibnamefont {Han}}, \bibinfo {author} {\bibfnamefont
  {Y.}~\bibnamefont {Matsui}}, \bibinfo {author} {\bibfnamefont
  {N.}~\bibnamefont {Nagaosa}}, \ and\ \bibinfo {author} {\bibfnamefont
  {Y.}~\bibnamefont {Tokura}},\ }\href {\doibase 10.1038/nature09124}
  {\bibfield  {journal} {\bibinfo  {journal} {Nature}\ }\textbf {\bibinfo
  {volume} {465}},\ \bibinfo {pages} {901} (\bibinfo {year}
  {2010})}\BibitemShut {NoStop}%
\bibitem [{\citenamefont {Seki}\ \emph
  {et~al.}(2012{\natexlab{a}})\citenamefont {Seki}, \citenamefont {Yu},
  \citenamefont {Ishiwata},\ and\ \citenamefont {Tokura}}]{Cu2OSeO3_skyrmions}%
  \BibitemOpen
  \bibfield  {author} {\bibinfo {author} {\bibfnamefont {S.}~\bibnamefont
  {Seki}}, \bibinfo {author} {\bibfnamefont {X.~Z.}\ \bibnamefont {Yu}},
  \bibinfo {author} {\bibfnamefont {S.}~\bibnamefont {Ishiwata}}, \ and\
  \bibinfo {author} {\bibfnamefont {Y.}~\bibnamefont {Tokura}},\ }\href
  {\doibase 10.1126/science.1214143} {\bibfield  {journal} {\bibinfo  {journal}
  {Science}\ }\textbf {\bibinfo {volume} {336}},\ \bibinfo {pages} {198}
  (\bibinfo {year} {2012}{\natexlab{a}})}\BibitemShut {NoStop}%
\bibitem [{\citenamefont {Yu}\ \emph {et~al.}(2012)\citenamefont {Yu},
  \citenamefont {Kanazawa}, \citenamefont {Zhang}, \citenamefont {Nagai},
  \citenamefont {Hara}, \citenamefont {Kimoto}, \citenamefont {Matsui},
  \citenamefont {Onose},\ and\ \citenamefont {Tokura}}]{yu2012}%
  \BibitemOpen
  \bibfield  {author} {\bibinfo {author} {\bibfnamefont {X.}~\bibnamefont
  {Yu}}, \bibinfo {author} {\bibfnamefont {N.}~\bibnamefont {Kanazawa}},
  \bibinfo {author} {\bibfnamefont {W.}~\bibnamefont {Zhang}}, \bibinfo
  {author} {\bibfnamefont {T.}~\bibnamefont {Nagai}}, \bibinfo {author}
  {\bibfnamefont {T.}~\bibnamefont {Hara}}, \bibinfo {author} {\bibfnamefont
  {K.}~\bibnamefont {Kimoto}}, \bibinfo {author} {\bibfnamefont
  {Y.}~\bibnamefont {Matsui}}, \bibinfo {author} {\bibfnamefont
  {Y.}~\bibnamefont {Onose}}, \ and\ \bibinfo {author} {\bibfnamefont
  {Y.}~\bibnamefont {Tokura}},\ }\href {\doibase 10.1038/ncomms1990} {\bibfield
   {journal} {\bibinfo  {journal} {Nature Commun.}\ }\textbf {\bibinfo {volume}
  {3}},\ \bibinfo {pages} {988} (\bibinfo {year} {2012})}\BibitemShut {NoStop}%
\bibitem{jonietz2010}
\bibinfo {author} {\bibfnamefont {F.}~\bibnamefont {Jonietz}},
\bibinfo {author} {\bibfnamefont {S.}~\bibnamefont {M{\"{u}}hlbauer}},
\bibinfo {author} {\bibfnamefont {C.}~\bibnamefont {Pfleiderer}},
\bibinfo {author} {\bibfnamefont {A.}~\bibnamefont {Neubauer}},
\bibinfo {author} {\bibfnamefont {W.}~\bibnamefont {M{\"{u}}nzer}},
\bibinfo {author} {\bibfnamefont {A.}~\bibnamefont {Bauer}},
\bibinfo {author} {\bibfnamefont {T.}~\bibnamefont {Adams}},
\bibinfo {author} {\bibfnamefont {R.}~\bibnamefont {Georgii}},
\bibinfo {author} {\bibfnamefont {P.}~\bibnamefont {B{\"{o}}ni}},
\bibinfo {author} {\bibfnamefont {R.~A.}~\bibnamefont {Duine}},
\bibinfo {author} {\bibfnamefont {K.}~\bibnamefont {Everschor}},
\bibinfo {author} {\bibfnamefont {M.}~\bibnamefont {Garst}},
\bibinfo {author} {\bibfnamefont {A.}~\bibnamefont {Rosch}},
\newblock {\bibinfo{journal}{Science}} \textbf{\bibinfo{volume}{330}},
  \bibinfo{pages}{1648--1651} (\bibinfo{year}{2010}).
\bibitem{fert2013}
\bibinfo{author}{A.~Fert}, \bibinfo{author}{V.~Cros}, and
  \bibinfo{author}{J.~Sampaio},
\newblock {\bibinfo{journal}{Nature Nanotech.}}
  \textbf{\bibinfo{volume}{8}}, \bibinfo{pages}{152--156}
  (\bibinfo{year}{2013}).

\bibitem [{\citenamefont {Ouyed}\ and\ \citenamefont
  {Butler}(1999)}]{Ouyed1999}%
  \BibitemOpen
  \bibfield  {author} {\bibinfo {author} {\bibfnamefont {R.}~\bibnamefont
  {Ouyed}}\ and\ \bibinfo {author} {\bibfnamefont {M.}~\bibnamefont {Butler}},\
  }\href@noop {} {\bibfield  {journal} {\bibinfo  {journal} {Astrophys.
  J.}\ }\textbf {\bibinfo {volume} {522}},\ \bibinfo {pages} {453}
  (\bibinfo {year} {1999})}\BibitemShut {NoStop}%
\bibitem [{\citenamefont {Luckock}\ and\ \citenamefont
  {Moss}(1986)}]{Luckock1986}%
  \BibitemOpen
  \bibfield  {author} {\bibinfo {author} {\bibfnamefont {H.}~\bibnamefont
  {Luckock}}\ and\ \bibinfo {author} {\bibfnamefont {I.}~\bibnamefont {Moss}},\
  }\href@noop {} {\bibfield  {journal} {\bibinfo  {journal} {Phys. Lett.
  B}\ }\textbf {\bibinfo {volume} {176}},\ \bibinfo {pages} {341} (\bibinfo
  {year} {1986})}\BibitemShut {NoStop}%
\bibitem{Lee2010}
\bibinfo{editor}{Brown, G.~E.} \& \bibinfo{editor}{Rho, M.} (eds.)
  \emph{\bibinfo{title}{The Multifaceted Skyrmion, Chap. 6}}
  (\bibinfo{publisher}{World Scientific}, \bibinfo{year}{2010}).
\bibitem [{\citenamefont {Castillejo}\ \emph {et~al.}(1989)\citenamefont
  {Castillejo}, \citenamefont {Jones}, \citenamefont {Jackson}, \citenamefont
  {Verbaarschot},\ and\ \citenamefont {Jackson}}]{Castillejo1989}%
  \BibitemOpen
  \bibfield  {author} {\bibinfo {author} {\bibfnamefont {L.}~\bibnamefont
  {Castillejo}}, \bibinfo {author} {\bibfnamefont {P.}~\bibnamefont {Jones}},
  \bibinfo {author} {\bibfnamefont {A.}~\bibnamefont {Jackson}}, \bibinfo
  {author} {\bibfnamefont {J.}~\bibnamefont {Verbaarschot}}, \ and\ \bibinfo
  {author} {\bibfnamefont {A.}~\bibnamefont {Jackson}},\ }\href@noop {}
  {\bibfield  {journal} {\bibinfo  {journal} {Nucl. Phys.}\ }\textbf {\bibinfo
  {volume} {A501}},\ \bibinfo {pages} {801} (\bibinfo {year}
  {1989})}\BibitemShut {NoStop}%
\bibitem [{\citenamefont {Kugler}\ and\ \citenamefont
  {Shtrikman}(1989)}]{Kugler1989}%
  \BibitemOpen
  \bibfield  {author} {\bibinfo {author} {\bibfnamefont {M.}~\bibnamefont
  {Kugler}}\ and\ \bibinfo {author} {\bibfnamefont {S.}~\bibnamefont
  {Shtrikman}},\ }\href {\doibase 10.1103/PhysRevD.40.3421} {\bibfield
  {journal} {\bibinfo  {journal} {Phys.Rev.}\ }\textbf {\bibinfo {volume}
  {D40}},\ \bibinfo {pages} {3421} (\bibinfo {year} {1989})}\BibitemShut
  {NoStop}%
\bibitem [{\citenamefont {Belesi}\ \emph {et~al.}(2012)\citenamefont {Belesi},
  \citenamefont {Rousochatzakis}, \citenamefont {Abid}, \citenamefont
  {R{\"{o}}{\ss}ler}, \citenamefont {Berger},\ and\ \citenamefont
  {Ansermet}}]{Cu2OSeO3_MEH}%
  \BibitemOpen
  \bibfield  {author} {\bibinfo {author} {\bibfnamefont {M.}~\bibnamefont
  {Belesi}}, \bibinfo {author} {\bibfnamefont {I.}~\bibnamefont
  {Rousochatzakis}}, \bibinfo {author} {\bibfnamefont {M.}~\bibnamefont
  {Abid}}, \bibinfo {author} {\bibfnamefont {U.~K.}\ \bibnamefont
  {R{\"{o}}{\ss}ler}}, \bibinfo {author} {\bibfnamefont {H.}~\bibnamefont
  {Berger}}, \ and\ \bibinfo {author} {\bibfnamefont {J.-{\mbox{Ph}}.}\
  \bibnamefont {Ansermet}},\ }\href {\doibase 10.1103/PhysRevB.85.224413}
  {\bibfield  {journal} {\bibinfo  {journal} {Phys. Rev. B}\ }\textbf {\bibinfo
  {volume} {85}},\ \bibinfo {pages} {224413} (\bibinfo {year}
  {2012})}\BibitemShut {NoStop}%
\bibitem [{\citenamefont {White}\ \emph {et~al.}(2012)\citenamefont {White},
  \citenamefont {Levatic}, \citenamefont {Omrani}, \citenamefont {Egetenmeyer},
  \citenamefont {Prsa}, \citenamefont {Zivkovic}, \citenamefont {Gavilano},
  \citenamefont {Kohlbrecher}, \citenamefont {Bartkowiak}, \citenamefont
  {Berger},\ and\ \citenamefont {Ronnow}}]{White}%
  \BibitemOpen
  \bibfield  {author} {\bibinfo {author} {\bibfnamefont {J.~S.}\ \bibnamefont
  {White}}, \bibinfo {author} {\bibfnamefont {I.}~\bibnamefont {Levatic}},
  \bibinfo {author} {\bibfnamefont {A.~A.}\ \bibnamefont {Omrani}}, \bibinfo
  {author} {\bibfnamefont {N.}~\bibnamefont {Egetenmeyer}}, \bibinfo {author}
  {\bibfnamefont {K.}~\bibnamefont {Prsa}}, \bibinfo {author} {\bibfnamefont
  {I.}~\bibnamefont {Zivkovic}}, \bibinfo {author} {\bibfnamefont {J.~L.}\
  \bibnamefont {Gavilano}}, \bibinfo {author} {\bibfnamefont {J.}~\bibnamefont
  {Kohlbrecher}}, \bibinfo {author} {\bibfnamefont {M.}~\bibnamefont
  {Bartkowiak}}, \bibinfo {author} {\bibfnamefont {H.}~\bibnamefont {Berger}},
  \ and\ \bibinfo {author} {\bibfnamefont {H.~M.}\ \bibnamefont {Ronnow}},\
  }\href@noop {} {\bibfield  {journal} {\bibinfo  {journal} {J. Phys.: Condens.
  Matter}\ }\textbf {\bibinfo {volume} {24}},\ \bibinfo {pages} {432201}
  (\bibinfo {year} {2012})}\BibitemShut {NoStop}%
\bibitem [{\citenamefont {Bos}\ \emph {et~al.}(2008)\citenamefont {Bos},
  \citenamefont {Colin},\ and\ \citenamefont {Palstra}}]{Bos08}%
  \BibitemOpen
  \bibfield  {author} {\bibinfo {author} {\bibfnamefont {J.-W.~G.}\
  \bibnamefont {Bos}}, \bibinfo {author} {\bibfnamefont {C.~V.}\ \bibnamefont
  {Colin}}, \ and\ \bibinfo {author} {\bibfnamefont {T.~T.~M.}\ \bibnamefont
  {Palstra}},\ }\href {\doibase 10.1103/PhysRevB.78.094416} {\bibfield
  {journal} {\bibinfo  {journal} {Phys. Rev. B}\ }\textbf {\bibinfo {volume}
  {78}},\ \bibinfo {pages} {094416} (\bibinfo {year} {2008})}\BibitemShut
  {NoStop}%
\bibitem [{\citenamefont {Belesi}\ \emph {et~al.}(2010)\citenamefont {Belesi},
  \citenamefont {Rousochatzakis}, \citenamefont {Wu}, \citenamefont {Berger},
  \citenamefont {Shvets}, \citenamefont {Mila},\ and\ \citenamefont
  {Ansermet}}]{Cu2OSeO3_NMR}%
  \BibitemOpen
  \bibfield  {author} {\bibinfo {author} {\bibfnamefont {M.}~\bibnamefont
  {Belesi}}, \bibinfo {author} {\bibfnamefont {I.}~\bibnamefont
  {Rousochatzakis}}, \bibinfo {author} {\bibfnamefont {H.~C.}\ \bibnamefont
  {Wu}}, \bibinfo {author} {\bibfnamefont {H.}~\bibnamefont {Berger}}, \bibinfo
  {author} {\bibfnamefont {I.~V.}\ \bibnamefont {Shvets}}, \bibinfo {author}
  {\bibfnamefont {F.}~\bibnamefont {Mila}}, \ and\ \bibinfo {author}
  {\bibfnamefont {J.~P.}\ \bibnamefont {Ansermet}},\ }\href@noop {} {\bibfield
  {journal} {\bibinfo  {journal} {Phys. Rev. B}\ }\textbf {\bibinfo {volume}
  {82}},\ \bibinfo {pages} {094422} (\bibinfo {year} {2010})}\BibitemShut
  {NoStop}%
\bibitem [{\citenamefont {Bogdanov}\ \emph {et~al.}(2002)\citenamefont
  {Bogdanov}, \citenamefont {R{\"o}{\ss}ler}, \citenamefont {Wolf},\ and\
  \citenamefont {M{\"u}ller}}]{bogdanov2002}%
  \BibitemOpen
  \bibfield  {author} {\bibinfo {author} {\bibfnamefont {A.~N.}\ \bibnamefont
  {Bogdanov}}, \bibinfo {author} {\bibfnamefont {U.~K.}\ \bibnamefont
  {R{\"o}{\ss}ler}}, \bibinfo {author} {\bibfnamefont {M.}~\bibnamefont
  {Wolf}}, \ and\ \bibinfo {author} {\bibfnamefont {K.-H.}\ \bibnamefont
  {M{\"u}ller}},\ }\href {\doibase 10.1103/PhysRevB.66.214410} {\bibfield
  {journal} {\bibinfo  {journal} {Phys. Rev. B}\ }\textbf {\bibinfo {volume}
  {66}},\ \bibinfo {pages} {214410} (\bibinfo {year} {2002})}\BibitemShut
  {NoStop}%
\bibitem [{\citenamefont {Seki}\ \emph
  {et~al.}(2012{\natexlab{b}})\citenamefont {Seki}, \citenamefont {Kim},
  \citenamefont {Inosov}, \citenamefont {Georgii}, \citenamefont {Keimer},
  \citenamefont {Ishiwata},\ and\ \citenamefont {Tokura}}]{Cu2OSeO3_NS}%
  \BibitemOpen
  \bibfield  {author} {\bibinfo {author} {\bibfnamefont {S.}~\bibnamefont
  {Seki}}, \bibinfo {author} {\bibfnamefont {J.-H.}\ \bibnamefont {Kim}},
  \bibinfo {author} {\bibfnamefont {D.~S.}\ \bibnamefont {Inosov}}, \bibinfo
  {author} {\bibfnamefont {R.}~\bibnamefont {Georgii}}, \bibinfo {author}
  {\bibfnamefont {B.}~\bibnamefont {Keimer}}, \bibinfo {author} {\bibfnamefont
  {S.}~\bibnamefont {Ishiwata}}, \ and\ \bibinfo {author} {\bibfnamefont
  {Y.}~\bibnamefont {Tokura}},\ }\href {\doibase 10.1103/PhysRevB.85.220406}
  {\bibfield  {journal} {\bibinfo  {journal} {Phys. Rev. B}\ }\textbf {\bibinfo
  {volume} {85}},\ \bibinfo {pages} {220406} (\bibinfo {year}
  {2012}{\natexlab{b}})}\BibitemShut {NoStop}%
\bibitem [{\citenamefont {Adams}\ \emph {et~al.}(2012)\citenamefont {Adams},
  \citenamefont {Chacon}, \citenamefont {Wagner}, \citenamefont {Bauer},
  \citenamefont {Brandl}, \citenamefont {Pedersen}, \citenamefont {Berger},
  \citenamefont {Lemmens},\ and\ \citenamefont
  {Pfleiderer}}]{Cu2OSeO3_ND_MH_single_cryst}%
  \BibitemOpen
  \bibfield  {author} {\bibinfo {author} {\bibfnamefont {T.}~\bibnamefont
  {Adams}}, \bibinfo {author} {\bibfnamefont {A.}~\bibnamefont {Chacon}},
  \bibinfo {author} {\bibfnamefont {M.}~\bibnamefont {Wagner}}, \bibinfo
  {author} {\bibfnamefont {A.}~\bibnamefont {Bauer}}, \bibinfo {author}
  {\bibfnamefont {G.}~\bibnamefont {Brandl}}, \bibinfo {author} {\bibfnamefont
  {B.}~\bibnamefont {Pedersen}}, \bibinfo {author} {\bibfnamefont
  {H.}~\bibnamefont {Berger}}, \bibinfo {author} {\bibfnamefont
  {P.}~\bibnamefont {Lemmens}}, \ and\ \bibinfo {author} {\bibfnamefont
  {C.}~\bibnamefont {Pfleiderer}},\ }\href {\doibase
  10.1103/PhysRevLett.108.237204} {\bibfield  {journal} {\bibinfo  {journal}
  {Phys. Rev. Lett.}\ }\textbf {\bibinfo {volume} {108}},\ \bibinfo {pages}
  {237204} (\bibinfo {year} {2012})}\BibitemShut {NoStop}%
\bibitem{Wilhelm2012}
\bibinfo {author} {\bibfnamefont {H.}~\bibnamefont {Wilhelm}},
\bibinfo {author} {\bibfnamefont {M.}~\bibnamefont {Baenitz}},
\bibinfo {author} {\bibfnamefont {M.}~\bibnamefont {Schmidt}},
\bibinfo {author} {\bibfnamefont {C.}~\bibnamefont {Naylor}},
\bibinfo {author} {\bibfnamefont {R.}~\bibnamefont {Lortz}},
\bibinfo {author} {\bibfnamefont {U.~K.}~\bibnamefont {R{\"o}{\ss}ler}},
\bibinfo {author} {\bibfnamefont {A.~A.}~\bibnamefont {Leonov}},
\bibinfo {author} {\bibfnamefont {A.~N.}~\bibnamefont {Bogdanov}},
\newblock {\bibinfo{journal}{J. Phys.: Condens. Matter}}
  \textbf{\bibinfo{volume}{24}}, \bibinfo{pages}{294204}
  (\bibinfo{year}{2012}).
\end{thebibliography}
\end{document}